\documentclass[useAMS,usenatbib]{mn2e}
\usepackage{epsfig}
\usepackage{graphicx}
\usepackage{times}
\usepackage{color}

\def\cm3{cm$^{-3}$}
\def\kms{km~s$^{-1}$}

\def\lsun{L$_{\odot}$}
\def\rsun{R$_{\odot}$}
\def\rstar{$R_{\star}$}

\def\msun{M$_{\odot}$}

\def\one{{\,\sc i}}
\def\two{{\,\sc ii}}
\def\three{{\,\sc iii}}

\def\beq{\begin{equation}}
\def\eeq{\end{equation}}
\def\rstar{$R_{\ast}$}
\def\zsun{Z$_{\odot}$}
\def\lesssim{\mathrel{\hbox{\rlap{\hbox{\lower4pt\hbox{$\sim$}}}\hbox{$<$}}}}
\def\gtrsim{\mathrel{\hbox{\rlap{\hbox{\lower4pt\hbox{$\sim$}}}\hbox{$>$}}}}

\def\cmfgen{{\sc cmfgen}}
\def\mesa{{\sc mesa star}}
\def\v1d{{\sc v1d}}

\newcommand{\iso}[2]{\ensuremath{^{#1}\rm{#2}}}

\def\aj{AJ}
\def\pasp{PASP}
\def\apj{ApJ}
\def\apjs{ApJS}
\def\apjl{ApJL}
\def\aap{A\&A}
\def\araa{ARA\&A}

\def\apss{Ap\&SS}
\def\mnras{MNRAS}
\def\nat{Nature}
\def\iaucirc{IAU~Circ.}

\def\ssr{Space Science Reviews}

\title[SNe II-P as metallicity probes]{Type II-Plateau supernovae as metallicity probes of the Universe}
\author[L. Dessart, C.~P. Gutierrez, M. Hamuy, D.~J. Hillier, T. Lanz et al.]
{L. Dessart,$^{1}$\thanks{This paper includes data gathered with the
6.5~m Magellan Telescopes located at Las Campanas Observatory, Chile; and
  the Gemini Observatory, Cerro Pachon,
  Chile (Gemini Program
  GS-2008B$-$Q$-$56). Based on observations collected at the European
  Organisation for Astronomical Research in the Southern Hemisphere,
  Chile (ESO Programmes 076.A-0156, 078.D-0048, 080.A-0516, and
  082.A-0526).}
C.~P. Gutierrez,$^{2}$
M. Hamuy,$^{2}$
D.~J. Hillier,$^{3}$
T. Lanz,$^{1}$
J.~P. Anderson,$^{2,4}$  \\ \\
{\LARGE \rm
G. Folatelli,$^{5}$
  W.~L. Freedman,$^{6}$
  F. Ley,$^2$
  N. Morrell,$^{7}$
    S.~E. Persson,$^{6}$
  M.~M. Phillips,$^{7}$
  } \\ \\
{\LARGE \rm
  M. Stritzinger,$^{8}$
  and N.~B. Suntzeff\,$^{9}$  } \\  \\
$^{1}$: Laboratoire Lagrange, UMR7293, Universit\'e Nice Sophia-Antipolis, CNRS,
Observatoire de la C\^{o}te dÕAzur, 06300 Nice, France. \\
$^{2}$:  Departamento de Astronom\'ia, Universidad de Chile, Casilla 36-D,
Santiago, Chile. \\
$^{3}$: Department of Physics and Astronomy \& Pittsburgh Particle physics, Astrophysics,
and Cosmology Center (PITT PACC), University of Pittsburgh,   \\
3941 O'Hara Street, Pittsburgh, PA 15260, USA. \\
$^{4}$: European Southern Observatory, Alonso de Cordova 3107, Vitacura, Santiago, Chile. \\
$^{5}$: Kavli Institute for the Physics and Mathematics of
  the Universe (WPI), Todai Institutes for Advanced Study, the
  University of Tokyo, Kashiwa, Japan 277-8583. \\
$^{6}$: Observatories of the Carnegie Institution of
  Washington, 813 Santa Barbara St., Pasadena, CA 91101, USA. \\
$^{7}$: Las Campanas Observatory, Carnegie Observatories,
  Casilla 601, La Serena, Chile. \\
$^{8}$: Department of Physics and Astronomy, Aarhus University,
Ny Munkegade 120, 8000 Aarhus C, Denmark. \\
$^{9}$: George P. and Cynthia Woods Mitchell Institute for
  Fundamental Physics and Astronomy, Department of Physics and
  Astronomy, Texas A\&M University, \\
  College Station, TX 77843, USA. \\
}

\date{Accepted 2014 February 27.  Received 2014 February 26; 
in original form 2014 January 25}

\pagerange{\pageref{firstpage}--\pageref{lastpage}} \pubyear{2014}

\begin{document}

\maketitle

\label{firstpage}

\begin{abstract}

We explore a method for metallicity determinations based on quantitative spectroscopy of
type II-Plateau (II-P) supernovae (SNe).
For consistency, we first evolve a set of 15\,\msun\ main sequence stars at 0.1, 0.4, 1,
and 2 $\times$ the solar metallicity. At the onset of core collapse, we trigger a piston-driven explosion
and model the resulting ejecta and radiation.
Our theoretical models of such red-supergiant-star explosions at different metallicity show that synthetic
spectra of SNe II-P possess optical signatures during the recombination phase that are sensitive to metallicity variations.
This sensitivity can be quantified and the metallicity inferred from the strengths of metal-line absorptions.
Furthermore, these signatures are not limited to O, but also include Na, Ca, Sc, Ti, or Fe.
When compared to a sample of SNe II-P from the Carnegie SN Project and previous
SN followup programs, we find that most events lie at a metallicity between 0.4 and 2$\times$ solar,
with a marked scarcity of SN II-P events at SMC metallicity.
This most likely reflects the paucity of low metallicity star forming regions in the local Universe.

SNe II-P have high plateau luminosities that make them observable spectroscopically at large distances.
Because they exhibit signatures of diverse metal species, in the future they may offer a means to
constrain the evolution of the composition (e.g., the O/Fe ratio) in the Universe out to a redshift of one
and beyond.

\end{abstract}

\begin{keywords}
radiative transfer -- supernovae: general
\end{keywords}

\section{Introduction}

Accurate metallicity measurements are desirable in numerous fields of astrophysics.
For example, mapping the evolution of metallicity with redshift is instrumental for understanding how
the composition of the Universe has evolved since Big Bang nucleosynthesis occurred.
Stars \citep{B2FH} and their supernovae (SNe; \citealt{arnett_96}) are the primary nuclear nurseries
and thus the metallicity evolution of the Universe allows us to look into the characteristics of the
stars and SNe that drove that evolution \citep{nomoto_etal_13}.
Accurate metallicity determinations are needed to interpret and to refine the mass metallicity
relation of galaxies \citep{tremonti_etal_04}, and to characterize the variation of metallicity
with galacto-centric radius in spiral galaxies, which is a probe of the star formation history
and galaxy dynamics \citep{boissier_prantzos_99}.
An additional motivation is the need to better understand
events that are expected theoretically to depend strongly on metallicity. A striking illustration
is the case of long-duration $\gamma$-ray bursts (GRBs), which are interpreted as stemming
from massive stars that avoid angular-momentum depletion through a metallicity-inhibited
stellar wind \citep{woosley_93,fruchter_etal_06}.

In distant astrophysical plasmas, the environmental metallicity is typically inferred
through the analysis of emission lines produced in photo-ionized nebulae \citep{osterbrock_89}.
While a variety of techniques are used within this approach, they tend to be limited
to oxygen abundance determinations \citep{kewley_dopita_02,pettini_pagel_04} and subject to
sizable systematic errors \citep{kewley_ellison_08}.
Other metal abundances, however, may not scale linearly with that of oxygen.
It is well known, for example, that oxygen is primarily produced by massive star
explosions \citep{woosley_weaver_95,arnett_96}, while SN Ia are the main contributors of iron-group
elements in the Universe. Hence, an interesting abundance ratio to seek is O/Fe to constrain their
relative contributions through the ages.

Furthermore, nebular line analyses can only be done in environments
where the gas density is high enough to produce detectable emission lines.
Star clusters in which the gas density is low, either intrinsically or following, for example, the action of massive
star winds, cannot be studied for their metallicity this way.
When applied to distant galaxies, this method suffers from irrevocable limitations in angular
resolution, yielding a metallicity averaged over an extended region rather than
the metallicity of a spatially restricted region.\footnote{This limitation can be somewhat overcome
for reasonably nearby SNe by using integral field spectroscopy (see, e.g., \citealt{rigault_etal_13}).}
This is problematic for determining the
metallicity at core-collapse SN sites \citep{anderson_etal_10,stoll_etal_13},
and in particular to quantify the metallicity bias between
standard SNe Ib/c sites and those of GRB/SNe \citep{modjaz_etal_08,modjaz_etal_11,sanders_etal_12}.

Modeling stellar spectra offers an alternative to nebular analyses. Quantitative spectroscopy of optical and/or
near-IR observations are used to determine the metal abundances in the star's atmosphere.
The advantage is that the theory of stellar atmospheres is well developed and accurate enough for
abundance determinations (see, e.g., \citealt{kudritzki_etal_12}). The drawback is that  stars are hard
to observe beyond the Local Group, limiting their use to the very nearby Universe.

In this paper, we present an attractive method for metallicity determination that is based on
SNe II-Plateau (II-P), which result from the core collapse and subsequent explosion
of red-supergiant (RSG) stars \citep{grassberg_etal_71,vandyk_etal_03,smartt_etal_04}.
With such ejecta, it is possible to overcome some of the limitations described above:
\begin{enumerate}
\item SNe II-P are very luminous and thus can be seen out to very large distances;
in the phase when the ejecta is optically thick, which lasts for $\sim$\,100\,d after explosion,
a standard SN II-P has a typical bolometric luminosity of a few times 10$^8$\,\lsun.
Although they are yet to be discovered (and proven to exist), SNe II-P resulting from the pair-production
instability are predicted to have even larger plateau luminosities
\citep{kasen_etal_11,dessart_etal_13},
hence even more attractive detection limits for transient surveys.
Over much of this high-luminosity phase,
their photosphere is at the hydrogen recombination temperature, i.e., $\sim$\,7000\,K.
Consequently, the bulk of this SN II-P radiation emerges in the optical.
It is thus particularly suited for observation of nearby and more distant events with optical
and near-IR instruments on (very/extremely) large telescopes.
\item Of all SNe, only SNe II-P are characterized by photospheres weakly affected
by either steady-state or explosive nuclear burning. Specifically, elements beyond O
(with the possible exception of Na) are unaffected by steady-state nuclear processing.
While chemical mixing may influence the
inner parts of the hydrogen-rich envelope of the progenitor RSG stars, the outer parts of the ejecta,
which are probed by the photosphere for up to $\sim$\,80\,d after explosion (see, e.g., Fig.~5 in
\citealt{DH11}) are essentially at the composition of the molecular cloud in which the progenitor star formed.
\item Photospheric phase SNe II-P spectra can be modelled with high fidelity using
non-LTE radiative-transfer codes like {\sc cmfgen} \citep{DH05_qs_SN, DH08_time,
HD12, dessart_etal_13b} or {\sc phoenix} \citep{baron_etal_07}.
Hence, quantitative spectroscopy can constrain the chemical composition at the photosphere,
as routinely done for stellar atmospheres (see, e.g., \citealt{HM98_lb,HM99_WC5}).
\item SNe II-P exhibit spectral
signatures associated with a variety of species, including intermediate-mass and iron-group elements.
For example, during the recombination epoch, SNe II-P show strong lines of H\one, O\one, Na\one, Ca\two,
Sc\two, Ti\two, and Fe\two\ (given in order of increasing atomic mass) --- this is a more diverse and eclectic
set than the [O\three]\,5007\,\AA\ and
[N\two]\,6584\,\AA\ lines seen in H\two\ regions and used for oxygen abundance determinations.
As we demonstrate in this paper, metallicity variations in the primordial composition
lead to distinct line strengths in the SN II-P optical spectrum as long as the SN photosphere probes
the outer progenitor envelope. Hence, in addition to oxygen, one can use a SN II-P spectrum to constrain
the mass fractions of additional species, including iron.
\end{enumerate}

The dependence of SN II-P spectra to metallicity variations has been discussed at a qualitative level
in the past for standard
RSG star explosions \citep{baron_etal_03,DH05_qs_SN,KW09,dessart_etal_13b}, as well as for
pair-instability SNe \citep{kasen_etal_11,dessart_etal_13}.  In this work, we use the
radiative-transfer simulations of \citet{dessart_etal_13b} to quantify the sensitivity
of a variety of SN II-P spectral signatures to metallicity variations, and investigate whether these
variations are sufficiently strong and well behaved to allow the inference of composition.
As this study is only meant as a proof of principle we limit the present discussion to
the restricted set of SN II-P models presented in \citet{dessart_etal_13b}, which correspond to the
explosion of a massive star evolved from a 15\,\msun\ main-sequence star with {\sc mesa}
\citep{paxton_etal_11,paxton_etal_13} at four different metallicities (one-tenth solar, two-fifth solar,
solar, and twice solar). In a future study we will cover a broader range of progenitor and explosion
properties, as well as metallicity values.

In the next section, we present the subset of models from \citet{dessart_etal_13b} that we
use in this work. We assess  the sensitivity of synthetic spectra to metallicity variations in
Section\,\ref{sect_res}, and present a preliminary  comparison to observations in
Section~\ref{sect_comp_obs}.
In Section~\ref{sect_conc}, we present our conclusions and discuss some interesting implications,
in particular the prospect of observing SNe II-P with very large and extremely large telescopes
to constrain the composition of the Universe out to cosmological distances.

\begin{table*}
\caption{
Summary of model properties used as initial conditions for \cmfgen\ simulations.
The first set of models are for different metallicities. The second set includes solar-metallicity
models evolved with different values of the mixing length parameter (models m15mlt2, also referred to as mlt2,
and m15z2m2 are the same). This second set is used merely to estimate systematic errors in Section~\ref{sect_res}.
}
\label{tab_progprop}
\begin{tabular}{l@{\hspace{2mm}}c@{\hspace{2mm}}c@{\hspace{2mm}}c@{\hspace{2mm}}
c@{\hspace{2mm}}c@{\hspace{2mm}}c@{\hspace{2mm}}c@{\hspace{2mm}}c@{\hspace{2mm}}c@{\hspace{2mm}}c@{\hspace{2mm}}
c@{\hspace{2mm}}c@{\hspace{2mm}}c@{\hspace{2mm}}c@{\hspace{2mm}}c@{\hspace{2mm}}}
\hline
   model   & $Z$   &    $M_{\rm final}$ & Age &  $T_{\rm eff}$  &    $R_{\star}$ &$L_{\star}$   &       $M_{\rm H}$
    &         $M_{\rm He}$  &   $M_{\rm O}$   &    $M_{r,{\rm Y_e}}$ &  $M_{r,\rm H}$ &  $M_{\rm H, env}$ &
          $M_{\rm ej}$   &  $E_{\rm kin}$  &  $M_{\iso{56}Ni}$  \\
          &     [\zsun]  &     [\msun]  & [Myr] &        [K]  &    [\rsun] &    [\lsun]      &   [\msun]   &  [\msun]  &    [\msun]
           &     [\msun]        &  [\msun]          &   [\msun]          & [\msun] & [B]  & [\msun]    \\
\hline
 m15z2m3   &    0.1  &  14.92   &  13.57   &    4144   &   524  &     72890  &    7.483 &      5.048  &     0.507     &   1.61 &      4.15  &     10.77  &    13.29  &   1.35  &   0.081  \\
 m15z8m3   &    0.4  &  14.76   &  13.34   &    3813   &   611  &     71052  &    7.183 &      5.252  &     0.428     &   1.63 &      4.09  &     10.67  &    13.12  &   1.27  &   0.036  \\
 m15z2m2   &    1.0  &  14.09   &  12.39   &    3303   &   768  &     63141  &    6.630 &      5.105  &     0.325     &   1.62 &      3.88  &     10.21  &    12.48  &   1.27  &   0.050  \\
 m15z4m2   &    2.0  &  12.60   &  10.88   &    3137   &   804  &     56412  &    5.119 &      5.042  &     0.387     &   1.40 &      3.77  &      8.83  &    11.12  &   1.24  &   0.095  \\
 \hline
m15mlt1   &     1.0  &  14.01   &  12.36   &    3318   &  1107  &    106958  &    6.516 &      5.167  &     0.354     &   1.36 &      3.89  &     10.13  &    12.57  &   1.24  &   0.121   \\
m15mlt2   &     1.0  &  14.09   &  12.39   &    3303   &   768  &     63141  &    6.630 &      5.105  &     0.325     &   1.62 &      3.88  &     10.21  &    12.48  &   1.27  &   0.050  \\
m15mlt3   &     1.0  &  14.08   &  12.41   &    4106   &   501  &     64218  &    6.542 &      5.173  &     0.383     &   1.54 &      3.92  &     10.16  &    12.52  &   1.34  &   0.086  \\
\hline
\end{tabular}
\flushleft
{\bf Notes:}
For the solar composition, we adopt the values of \citet{GS98} and scale each metal mass fraction
as indicated. Following the columns containing the cumulative ejecta masses in H, He, and O, we give the
Lagrangian mass corresponding to specific locations in the pre-SN star. We give
$M_{r,{\rm Y_e}}$, which corresponds to the edge of the iron core, i.e.,  where the electron fraction $Y_e$ rises
to 0.49 as we progress outward from the star center (this is also the piston location for the explosion), and
$M_{r,\rm H}$ which corresponds to the base of the H-rich envelope (i.e., the helium-core mass).
The last three columns give some ejecta properties that result from the imposed explosion parameters.
The quoted  \iso{56}Ni mass corresponds to that originally produced in the explosion.
\end{table*}

\begin{figure*}
\epsfig{file=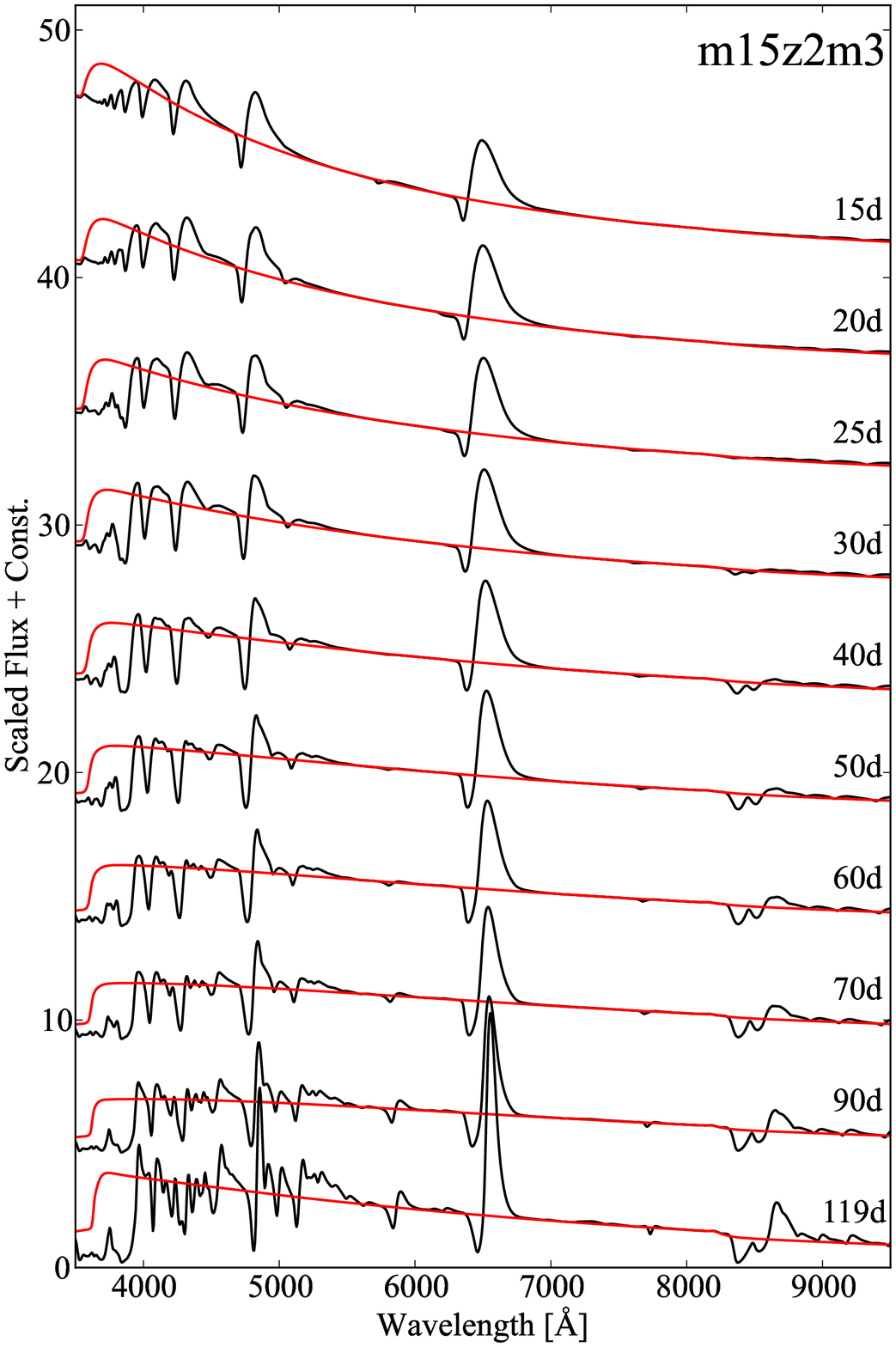,width=8.75cm}
\epsfig{file=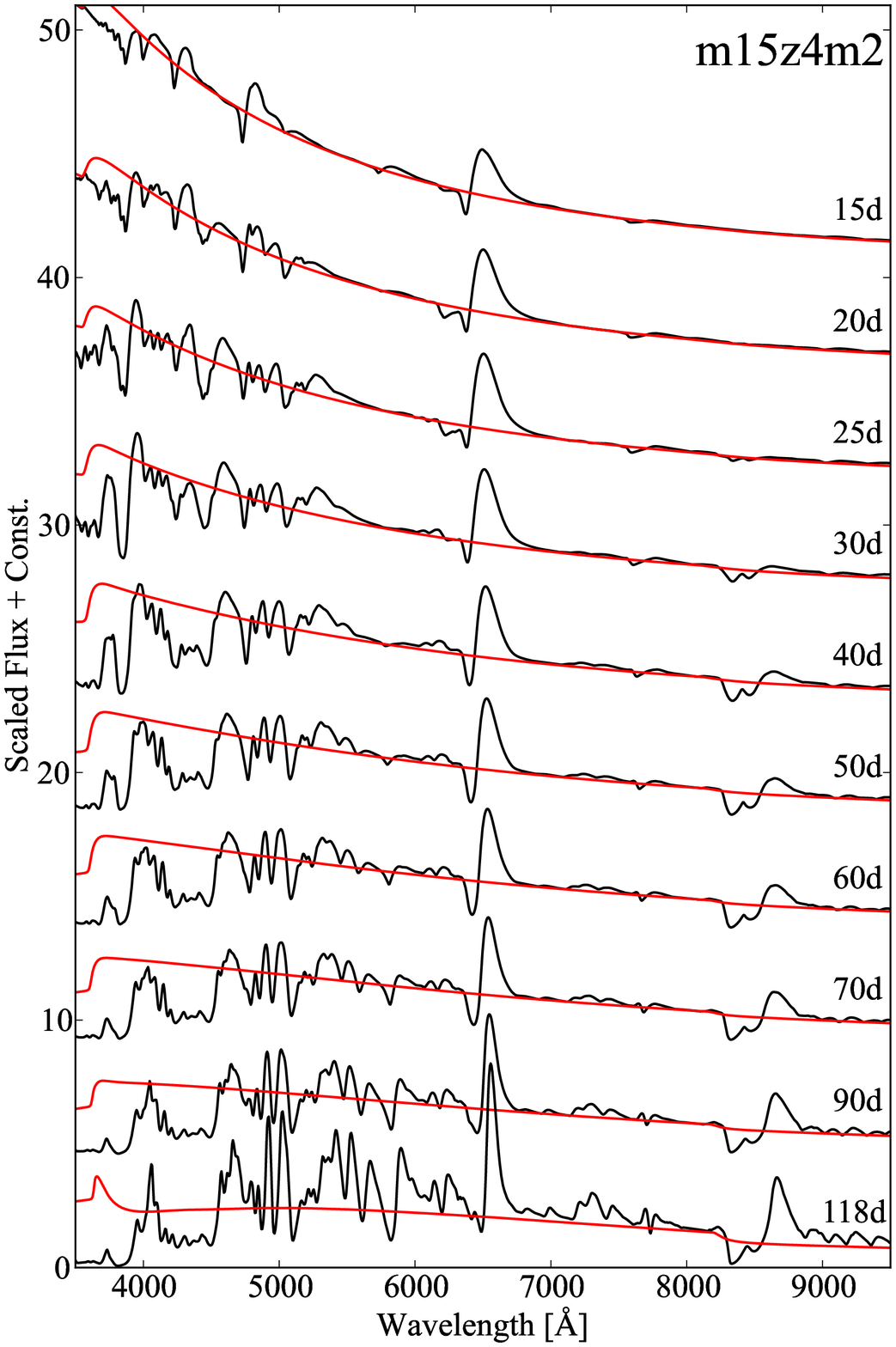,width=8.75cm}
\caption{Spectral evolution in the optical range for SN II-P models at one-tenth (m15z2m3; left)
and twice (m15z4m2; right) the solar metallicity, highlighting their sensitivity to metallicity variations.
The UV range is also sensitive to metallicity variations but lines tend to overlap and to be saturated,
which complicates the analysis. The red curve corresponds to the synthetic continuum flux.
Line identifications are discussed in detail in the appendix of \citet{dessart_etal_13b}.
\label{fig_spec_seq}
}
\end{figure*}

\section{Summary of numerical setup}
\label{sect_setup}

   The simulations we discuss in this work have been presented in \citet{dessart_etal_13b}.
For completeness, we summarize the numerical procedure for our calculations.

  Using \mesa, \citet{dessart_etal_13b} generated a grid of models starting with the same main-sequence
mass of 15\,\msun\ but varying a number of parameters known to influence stellar evolution.
In this paper, we focus on the influence of metallicity variations, and present results for cases
of 0.002, 0.008, 0.02 (taken as our solar metallicity), and  0.04 (named m15z2m3, m15z8m3, m15z2m2, and m15z4m2).
When varying the metallicity, we merely scale the mass fraction of each metal
by a factor 1/10, 2/5, 1, and 2.
The \mesa\ simulations are performed adopting zero rotation, a mixing-length parameter $\alpha=1.6$
(we adopt the Schwarschild criterion for convection),
a standard resolution ({\sc mesh\_delta\_coeff}=1), no core-overshooting, the mass loss rate recipes dubbed ``Dutch''
with a scaling of 0.8.
At the end of its life, the solar metallicity model m15z2m2 is a 14.09\,\msun\ RSG star with a luminosity of 63141\,\lsun,
a radius of 768\,\rsun, and an effective temperature of 3303\,K. It possesses an H-rich envelope of 10.21\,\msun\ and
an helium core of 3.88\,\msun\ (set equal to the Lagrangian mass at the inner edge of the H-rich envelope).
The outer edge of the iron core is at a Lagrangian mass of 1.6\,\msun.

Because of the adopted metallicity dependence of RSG mass loss rates, our \mesa\ models
m15z2m3, m15z8m3, m15z2m2, and m15z4m2 have a final H-rich envelope mass that depends
on Z, ranging from 10.77 (m15z2m3) down to 8.83\,\msun\ (m15z4m2).
The efficiency of convective energy transport being set in all four
simulations through a mixing-length parameter of 1.6, the variation in
metallicity, which changes the opacity in the envelope, alters the stellar
radius (since the energy flux to transport from the edge of the core to the stellar surface is
essentially the same between these 4 models). Consequently, for smaller metallicities (opacities), we
obtain smaller RSG radii, with values between 524\,\rsun\ (m15z2m3) and 804\,\rsun\ (m15z4m2).
Our RSG models are thus both more massive and more compact at lower metallicities.
It is important to note that all our models encounter core collapse in the RSG phase. Hence, at
least from a theoretical standpoint, there is no reason to believe that 15\,\msun\ stars would not yield
SNe II-P over a large range of metallicities.

\begin{figure*}
\epsfig{file=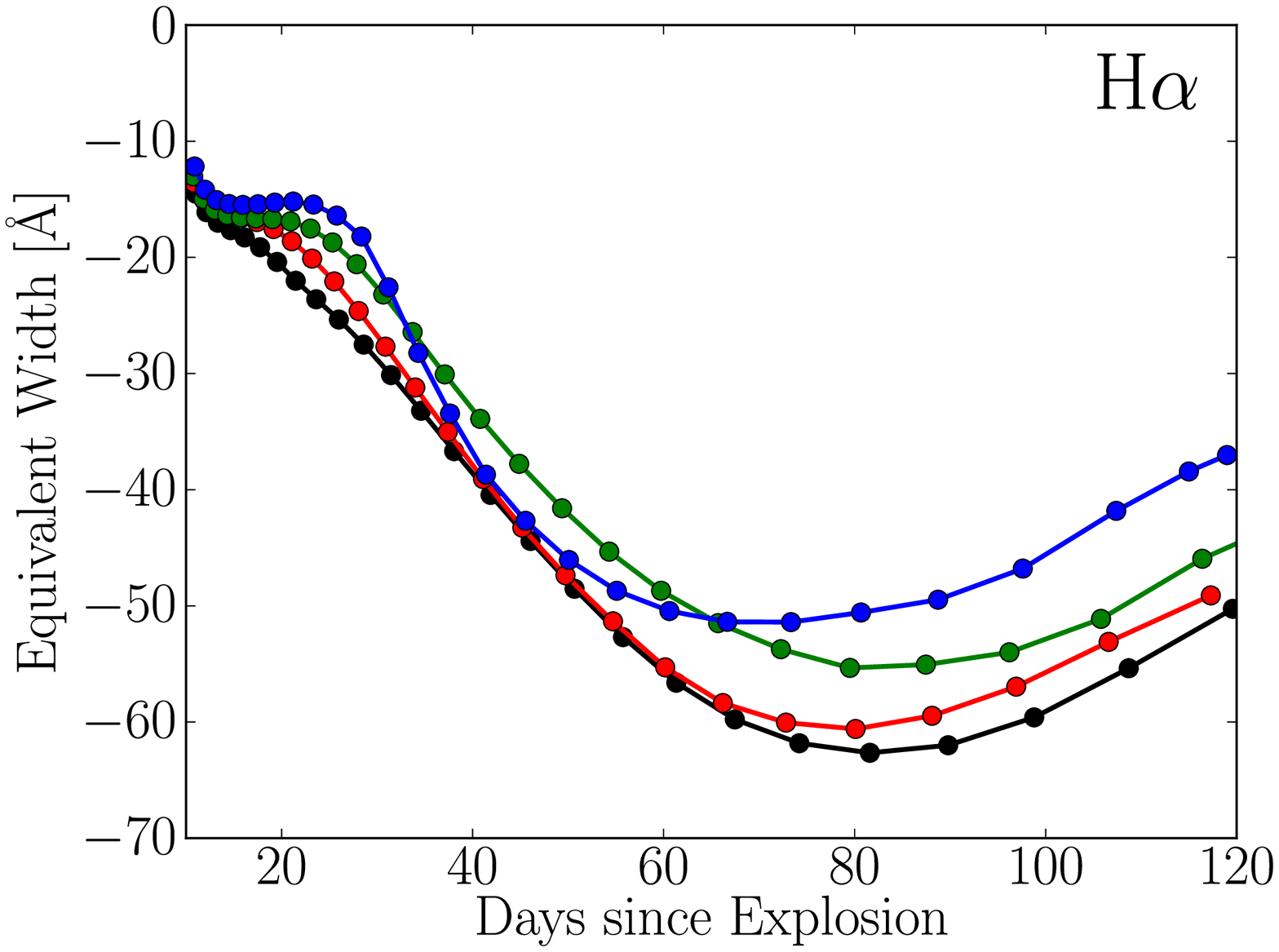,width=8.75cm}
\epsfig{file=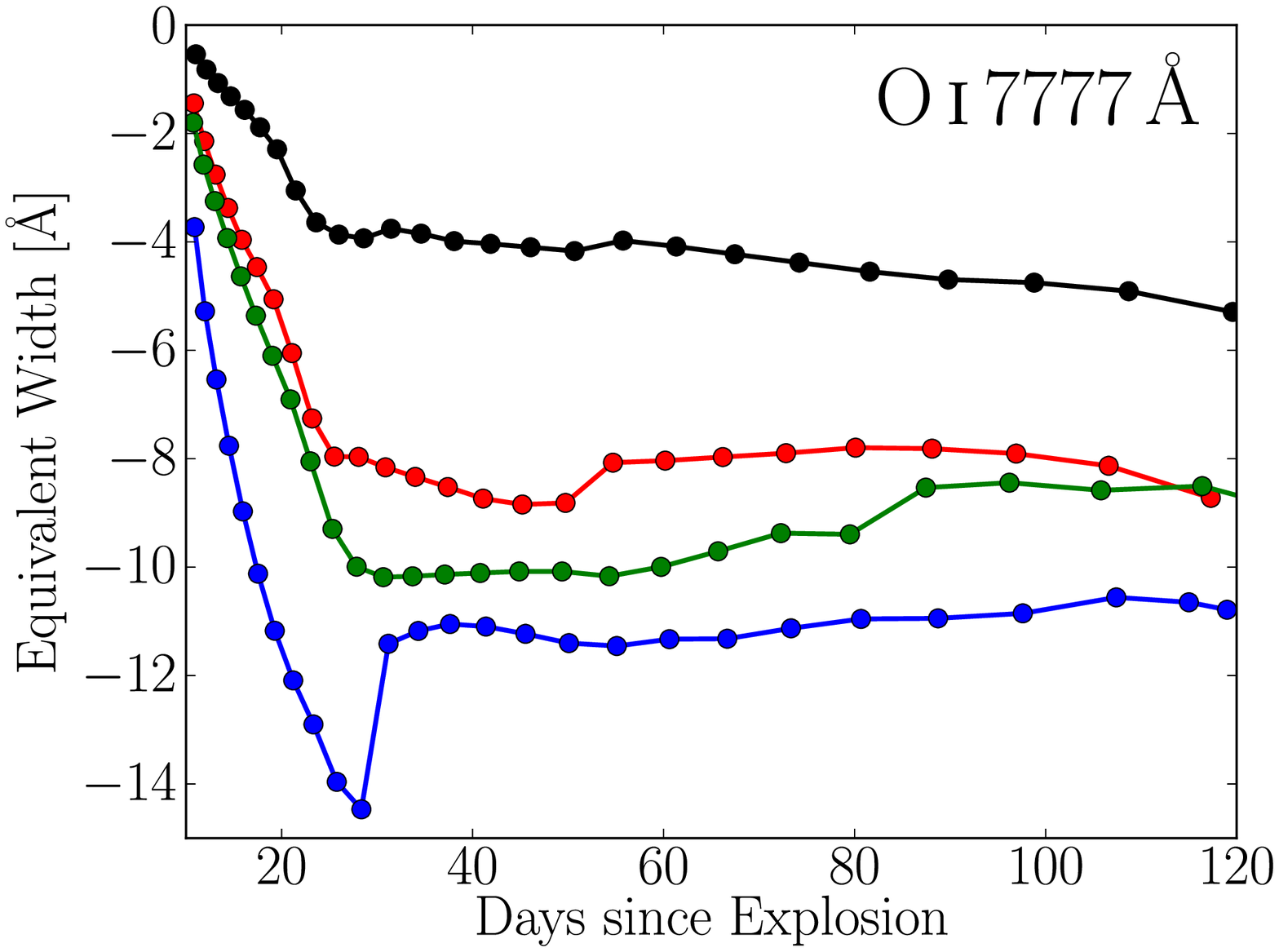,width=8.75cm}
\epsfig{file=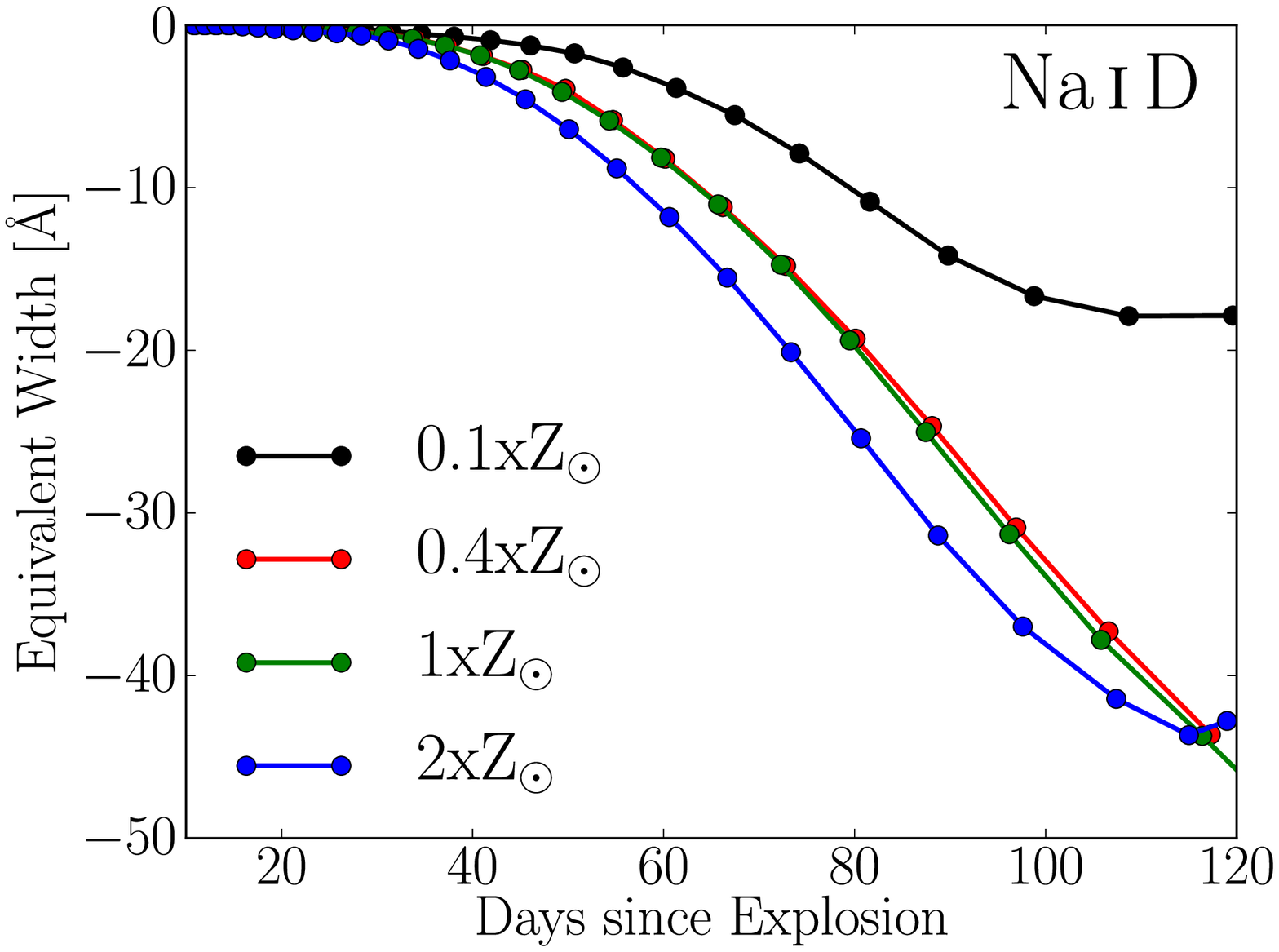,width=8.75cm}
\epsfig{file=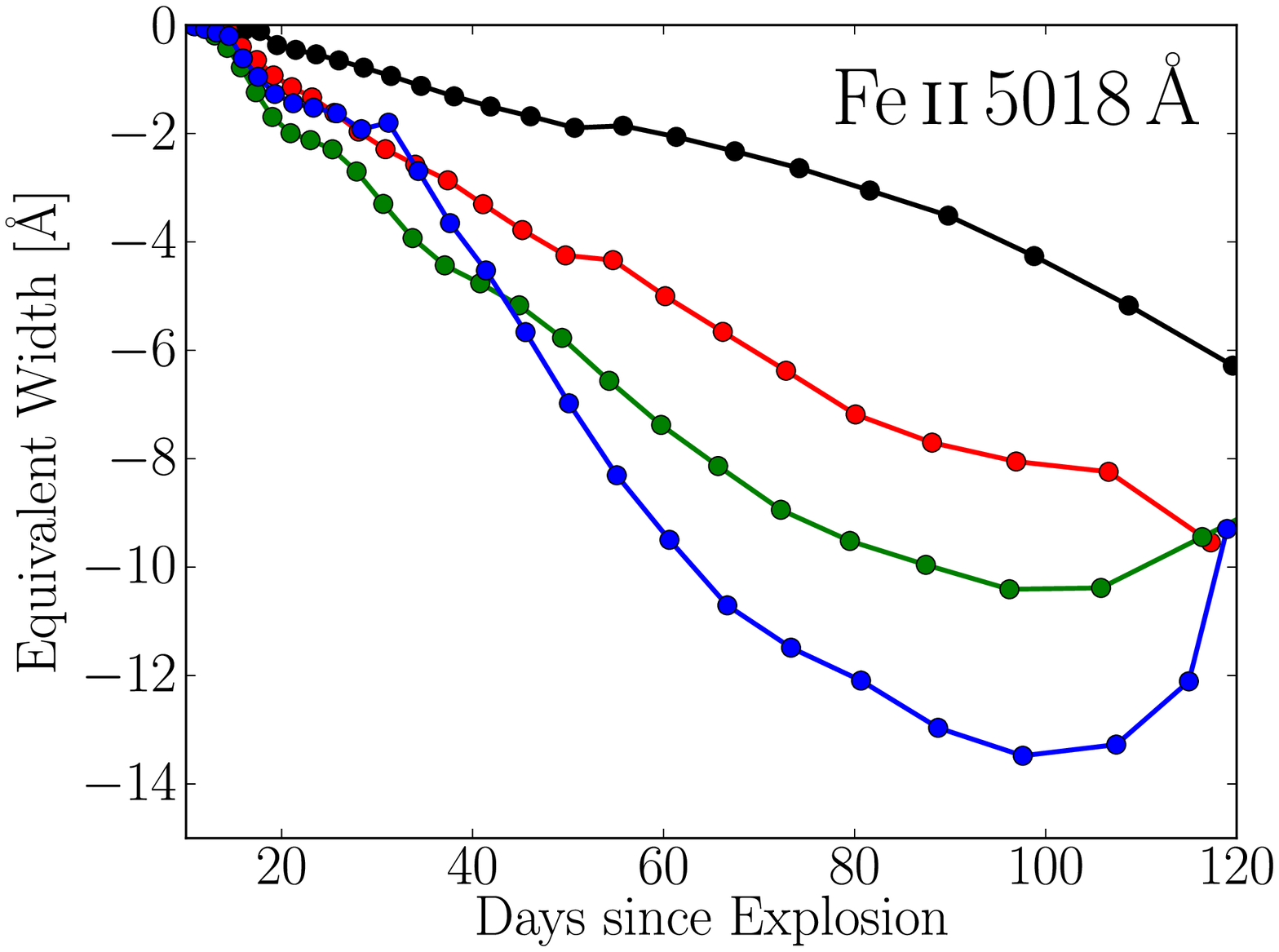,width=8.75cm}
\epsfig{file=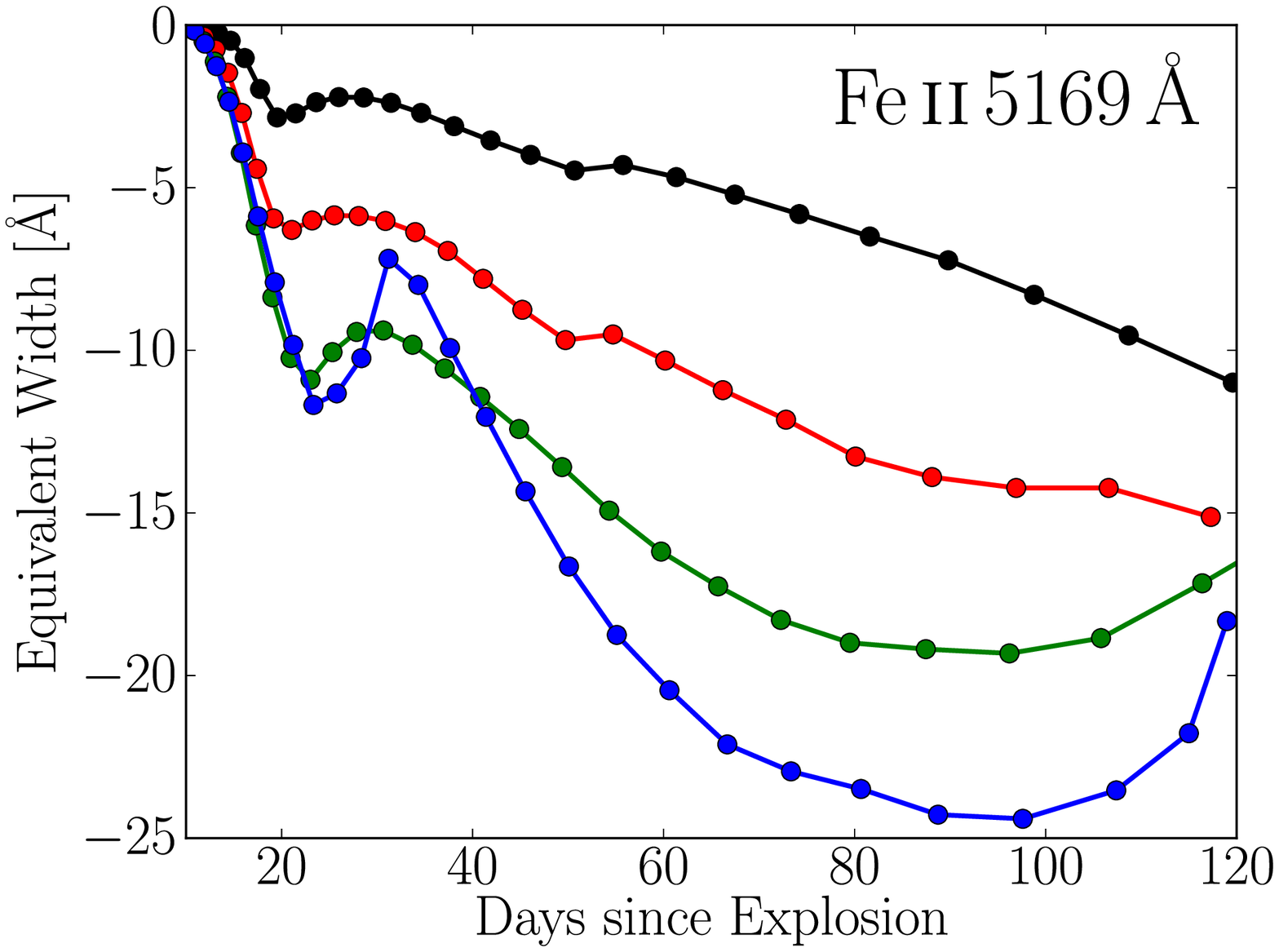,width=8.75cm}
\epsfig{file=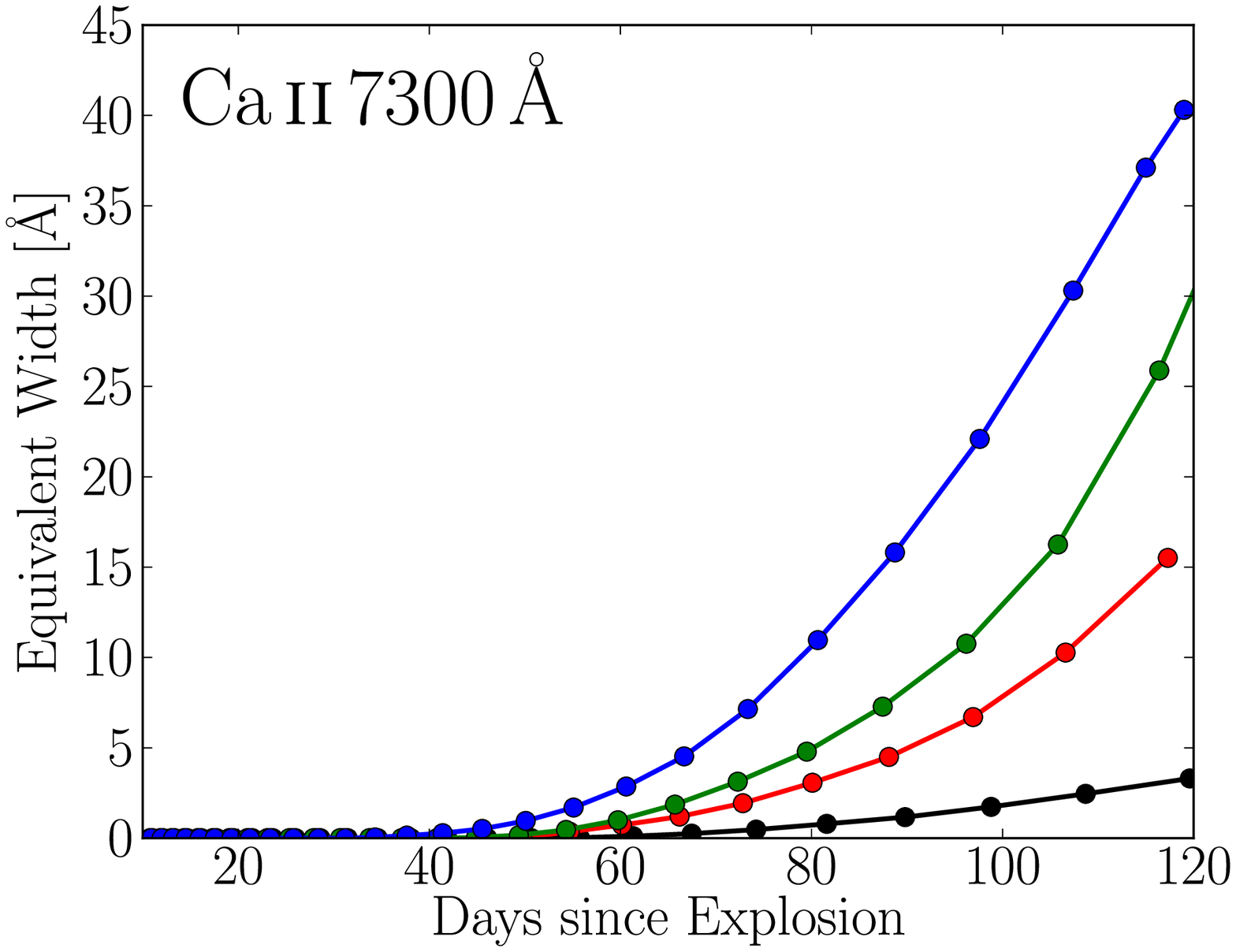,width=8.75cm}
\caption{{\bf First five panels from top left:}
Evolution of the line-absorption EW associated with H$\alpha$ (top left), O\one\,7777\,\AA\ (top right),
Na\one\,D (middle left), Fe\two\,5018\,\AA\ (middle right), and Fe\two\,5169\,\AA\ (bottom left).
{\bf Bottom-right panel:} Evolution of the EW for the pure emission feature associated with the
Ca\two\,7300\,\AA\ doublet.
Each colored curve corresponds to a different ejecta model resulting from the explosion of a 15\,\msun\ star
evolved at a different metallicity (see Section~\ref{sect_setup} for details).
\label{fig_ew}
}
\end{figure*}

When a mass cell within the iron-core reaches an infall velocity of 100\,\kms, each \mesa\ simulation is stopped and
a piston-driven explosion is simulated with the grey radiation-hydrodynamics code \v1d\ \citep{livne_93}. Following
the procedure outlined in \citet{DLW10b}, the piston deposits an energy equal to the binding energy of the envelope
plus the asymptotic kinetic energy at infinity, which we select to be 1.2\,B.
\v1d\ treats explosive nucleosynthesis so the change in composition following shock passage is computed.
In piston-driven explosions, the amount of \iso{56}Ni produced depends sensitively on the adopted location
of the piston and the timescale over which it injects energy. For the current set of simulations, the piston was
placed at the base of the Si-rich shell and was given a constant velocity of 10000\,\kms\ until the desired energy
was reached.
In this artificial approach, \iso{56}Ni is only produced in the densest (i.e., innermost) ejecta regions crossed by the piston.
It is thus essential to apply an artificial mixing to these 1-D simulations in order to mimic the mixing that
takes place in more realistic multi-D simulations of core-collapse SNe \citep{hammer_etal_10}.

 The mixing we apply is moderate. For example, the initial mass fraction of \iso{56}Ni
in the four ejecta modeled here is $\sim$10$^{-6}$ at 2500\,\kms\
(a location reached by the photosphere at $\sim$\,100\,d after explosion; see also Fig.~5 of \citealt{DH11}).
Hence, in our models, radioactive decay has a negligible
influence on the SN II-P radiation for most of the plateau length, either through heating,
non-thermal ionization/excitation, or abundance variations.
In Nature, mixing may be stronger and may influence the radiation earlier on. To ensure we probe only
the initial star metallicity, it will be necessary to focus on the mid-plateau phase.
In higher energy SN II-P explosions, more \iso{56}Ni may be produced and the plateau may be shorter
and/or slanted (see Section\,9 of \citealt{dessart_etal_13b}).
It is preferable to focus on standard energy explosions, with properties comparable to those of, e.g., SN\,1999em.

About ten days after shock breakout, homologous expansion is reached. The full ejecta is remapped into
\cmfgen\ and evolved until nebular times. A summary of pre-SN and ejecta properties as well as model
light curves are given in \citet{dessart_etal_13b}. These four ejecta have a similar color evolution, plateau
length etc. but they show striking differences in spectral lines, which appear to reflect the variation in the
metallicity of their progenitors. The focus of our present study is to discuss these metallicity signatures.

\section{Synthetic signatures of metallicity variations in SN\lowercase{e} II-P}
\label{sect_res}

   Discussing spectroscopic differences within a set of SN II-P models is complicated
   by the potentially distinct rates of evolution of their color and ejecta ionization. To
   disentangle abundance and ionization effects, which are both known to alter line strengths,
   it is best to compare these SN II-P models during the photospheric phase and when they
   have the same color, as done in \citet{dessart_etal_13b} at  $U-V$ of $-$0.4, 0.4, and 1.7\,mag.
    Despite the comparable spectral energy distributions at a given color, our set of SN II-P models
    shows considerable diversity in individual lines strengths, and in particular during the recombination
    phase when metal-line blanketing strengthens. This phase typically starts a month after explosion, although
    variations in progenitor radius can delay its onset \citep{dessart_etal_13b}.
    The lower the metallicity, the weaker the metal lines appear (Fig.~\ref{fig_spec_seq}).
    The most obvious variations are in the following spectral lines or groups of lines (given in order
    of increasing atomic mass of the parent species):

\begin{itemize}
    \item O\one\,7777\,\AA,
    \item Na\one\,D,
    \item  the Ca\,\two\ triplet at 8500\,\AA\ and the semi-forbidden-line doublet at 7300\,\AA,
    \item the Ti\two\ broad blanketing region ranging from 4200 to 4500\,\AA, and associated
    with the atomic configurations 3d$^2$4s---3d$^2$4p, 3d$^3$---3d$^2$4p,
    \item Sc\two\  5239\,\AA\ (4s$^2$ \,$^1$Se---4p\,$^1$Po), 5526\,\AA\ (3d$^2$\,$^1$Ge---4p\,$^1$Fo),
    5669\,\AA\ (3d$^2$\,$^1$GPe---4p\,$^3$Po), and companion configurations,
    \item Fe\,\two\ 4923, 5018, and 5169\,\AA\ (3d$^5$4s$^2$Se---3d$^6$\,4p\,Po terms).
\end{itemize}

Because of line overlap, numerous other lines, from these or other metal species, prevent a clear connection to
a metallicity variation, particularly when the metallicity is super solar (model m15z4m2).
In this case, the total flux (black curve) departs significantly from the continuum flux level (red curve), even
between neighboring P-Cygni profiles. In contrast, even when metal-line blanketing is
strong, the lower metallicity model m15z2m3 still exhibits well isolated P-Cygni profiles,
with the flux going back to the continuum level in-between neighboring features.

The UV range is very sensitive to the metal composition at the photosphere. Indeed,
it is strongly attenuated by metal-line blanketing, even at one-tenth solar
metallicity. Numerous metal lines overlap and may remain saturated even when the metal abundance decreases.
Bumps and valleys in the UV range do not correspond to specific line features, but reflect
absorption variations instead. So, bumps are merely regions of reduced absorptions
(i.e., not residual emission above the continuum level).
Hence, one cannot in general associate a given feature with a specific metal line, complicating the inference
of metal abundances from UV spectra.
The UV flux is also faint because of the low temperature of the photosphere during the recombination epoch.
Hence, the UV domain is not ideal if we wish to determine robust metallicity signatures from SNe II-P spectra over
a large range of distances.

The strong differences shown in Fig.~\ref{fig_spec_seq} can be quantified by means of line-absorption equivalent
widths (EWs), which we measure for each time step in each model sequence.
With \cmfgen, we can compute the continuum flux for each epoch and thus calculate the true EW  for any feature.
However, line overlap often prevents the association of a given feature with a given transition in a given atom/ion.
To identify an abundance effect on specific lines, which is our first goal, we compute the spectrum for each species individually.
Doing these EW calculations on theoretical models is automatic and unambiguous, which is why
in this model section we show such measurements, as opposed to the standard pseudo-equivalent
width (pEW)  extracted from observed spectra.
We show the results of true EW calculations on H$\alpha$, O\one\,7777\,\AA, Na\one\,D,
Fe\two\,5018\,\AA, Fe\two\,5169\,\AA, and Ca\two\,7300\,\AA\ in Fig.~\ref{fig_ew}.

The H$\alpha$ EW becomes more and more negative as the SN model proceeds through the photospheric
phase. This is not a signature of an abundance variation but simply a reflection of the changing density and
temperature conditions in the spectrum formation region \citep{DH11,dessart_etal_13b}, as well as the
increasing importance of time-dependent effects on the ionization \citep{UC05,DH08_time}.
The small differences in H$\alpha$ EW among sequences are caused by small variations in blanketing
of the UV flux. This affects the radiation through the photosphere and the formation of H$\alpha$,
but the change in metallicity leaves the H mass fraction basically untouched.

A much stronger trend of increasing absorption/emission is present for lines associated with under-abundant species
(Fig.~\ref{fig_ew}). The magnitude of the effect varies considerably from low to high metallicity, which
is not surprising since metallicity variations yield significant changes in the associated species mass fraction.
The evolution of the EW can be non-monotonic, and its magnitude may behave non-linearly with metallicity.
If we select mid-plateau epochs ($\sim$\,50\,d after explosion),
we see that the O and Fe line strengths reflect with fidelity the variations in metallicity.
The effect is strong --- between models m15z2m3 and m15z4m2, the EW of
O\,\one\,7777\,\AA\ and Fe\,\two\ 5018\,\AA\ or Fe\,\two\ 5169\,\AA\ varies by a factor of 3-4.

The Ca\two\ semi forbidden lines at 7300\,\AA, which are optically thin and in emission,
also show a strong sensitivity to metallicity. These
lines are observed and predicted here in our models during the second half of the plateau phase, hence
well before the nebular phase. The associated line emission takes place in the
outer SN ejecta (the former H-rich envelope) whose density is much lower than that of the inner ejecta
at those epochs. In the SN context, the ejecta temperature and electron density are well constrained by the model
and are comparable for the four models discussed here. Thus, the differing line strengths
stem exclusively from variations in the Ca abundance in the H-rich envelope and hence offer an additional
probe of the metallicity.
The differences between models are huge (i.e., a factor of 10 between models m15z2m3 and m15z4m2).
An important signature is that the model with a metallicity of one-tenth solar shows negligible
Ca\two\,7300\,\AA\ emission during the photospheric phase.

\begin{figure}
\epsfig{file=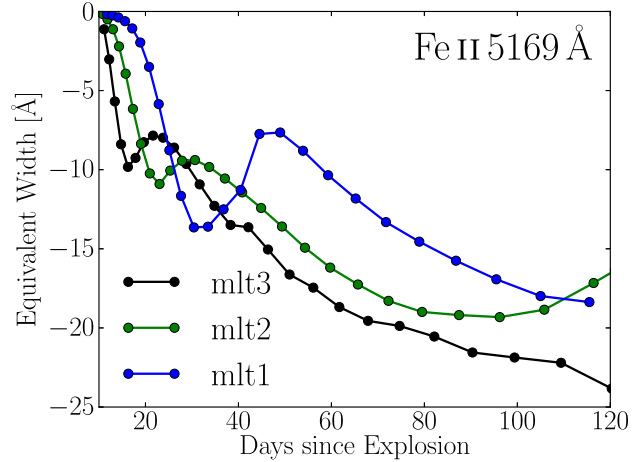,width=8.85cm}
\caption{
Evolution of the EW in Fe\two\,5169\,\AA\ for solar-metallicity models m15mlt1, m15mlt2, m15mlt3
(labels omit ``m15'' since redundant).
Here, the distinct tracks stem from the variation in progenitor radius for the three progenitor star models.
To reduce systematic errors when inferring metallicities, it is important to employ a SN II-P model
that corresponds closely to the SN under study.
\label{fig_sys_mod}
}
\end{figure}

While there is no sizable error on the EW measurements performed on synthetic spectra,
there are systematic errors associated with the modeling and these are hard to estimate. A lot of
work has been devoted to test the code and ensure the results are as accurate as possible. We use
a set of large model atoms, in particular for Fe\one\ and Fe\two, because it is essential at the recombination epoch
to obtain accurate colors \citep{dessart_etal_13b}. In our simulations, doubling the spatial or frequency resolution
does not change the synthetic spectra. The results are apparently well converged in a numerical sense.

The main source of error in assessing metallicities from synthetic spectra will likely come from using
ejecta or progenitor star models that are not suitable for the observed SN under consideration.
For example, progenitor stars of different surface radii at the time of
explosion produce SNe II-P that follow a different color evolution, exhibiting a significant
variation in the time at which recombination occurs.
\citet{dessart_etal_13b} explored this issue by means of physical stellar evolution models
(at solar metallicity) in which the mixing-length describing convection was modified from the standard value of 1.6
(model mlt2; \rstar$=$\,768\,\rsun)
to 1 (model mlt1;  \rstar$=$\,1107\,\rsun) and 3 (model mlt3;  \rstar$=$\,501\,\rsun). Only the model mlt3 fitted
the observations of SN1999em, implying the stellar radius is not a free parameter but is instead
constrained from observations --- the other models remained too blue for too long.
In Fig.~\ref{fig_sys_mod}, we show the evolution of the EW in Fe\two\,5169\,\AA\ for these three models.
The different times for the onset of recombination is evident; the models that recombine first (more compact progenitor),
also show the strongest EW at the recombination epoch.
The differences here are tied to the progenitor radius, while the metallicity, being solar for
all three models, plays no role.
This illustration gives some measure of the potential errors in our method but it gives a clear overestimate.
When modeling a SN, the progenitor radius  is one entity that we constrain ---
it is not a free parameter, and the same holds for the envelope mass or the explosion energy.
Hence, it is important to select carefully the SN II-P we employ for metallicity determinations.
As we discuss below, it is critical to exclude from the present study
all SNe II-P with a significant slant in $V$-band light
curve during the photospheric phase, because unlike our present models they are not
genuine plateau SNe. Furthermore, in the present study, we must select observations that are roughly compatible with
our four models m15z2m3, m15z8m3, m15z2m2, and m15z4m2 (similar kinetic energy, similar color evolution). In the future,
we will need to produce tailored models for each observed SN under study in order to limit potential systematic errors.

\begin{figure*}
\epsfig{file=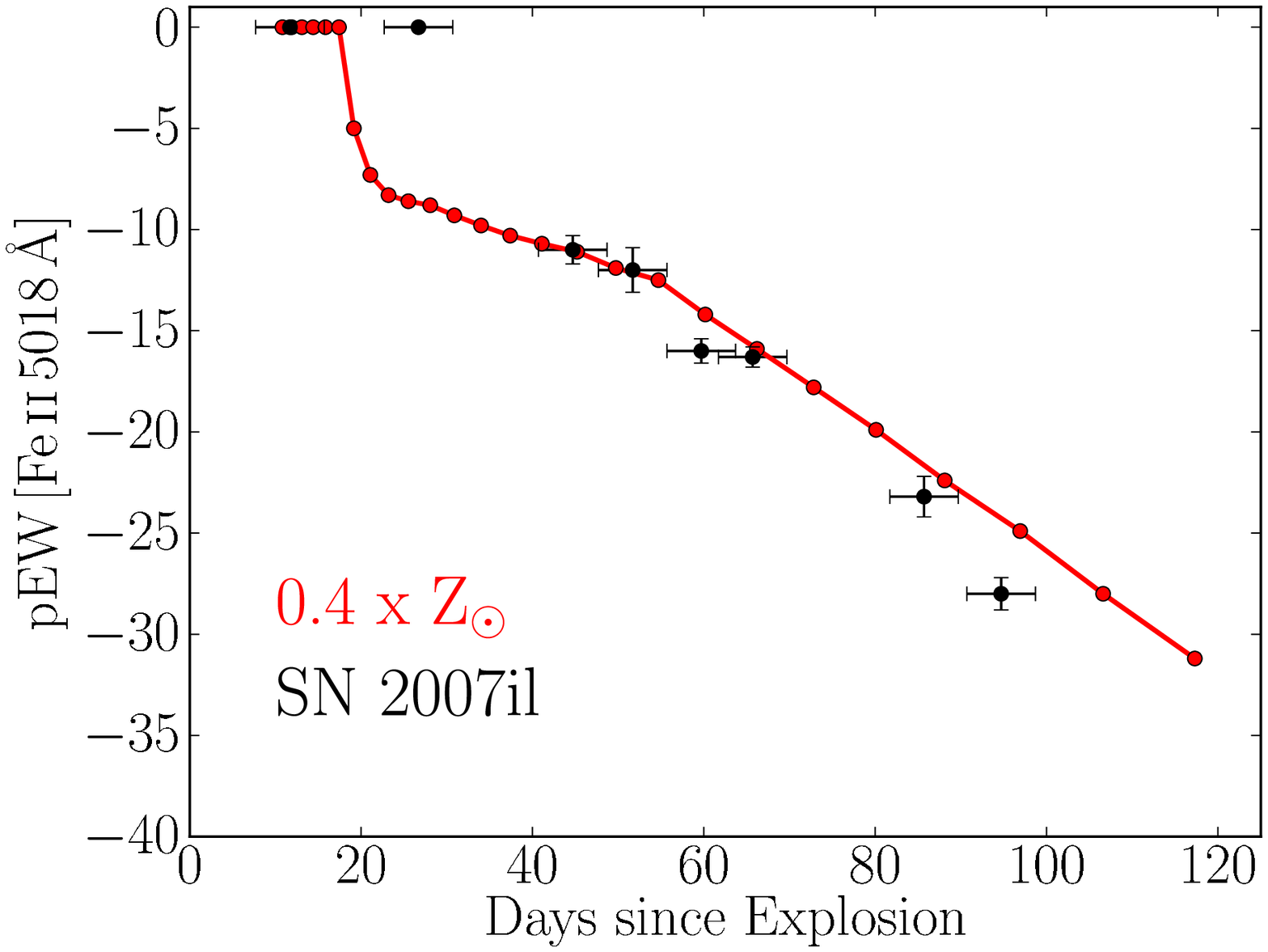,width=5.75cm}
\hspace{0.1cm}
\epsfig{file=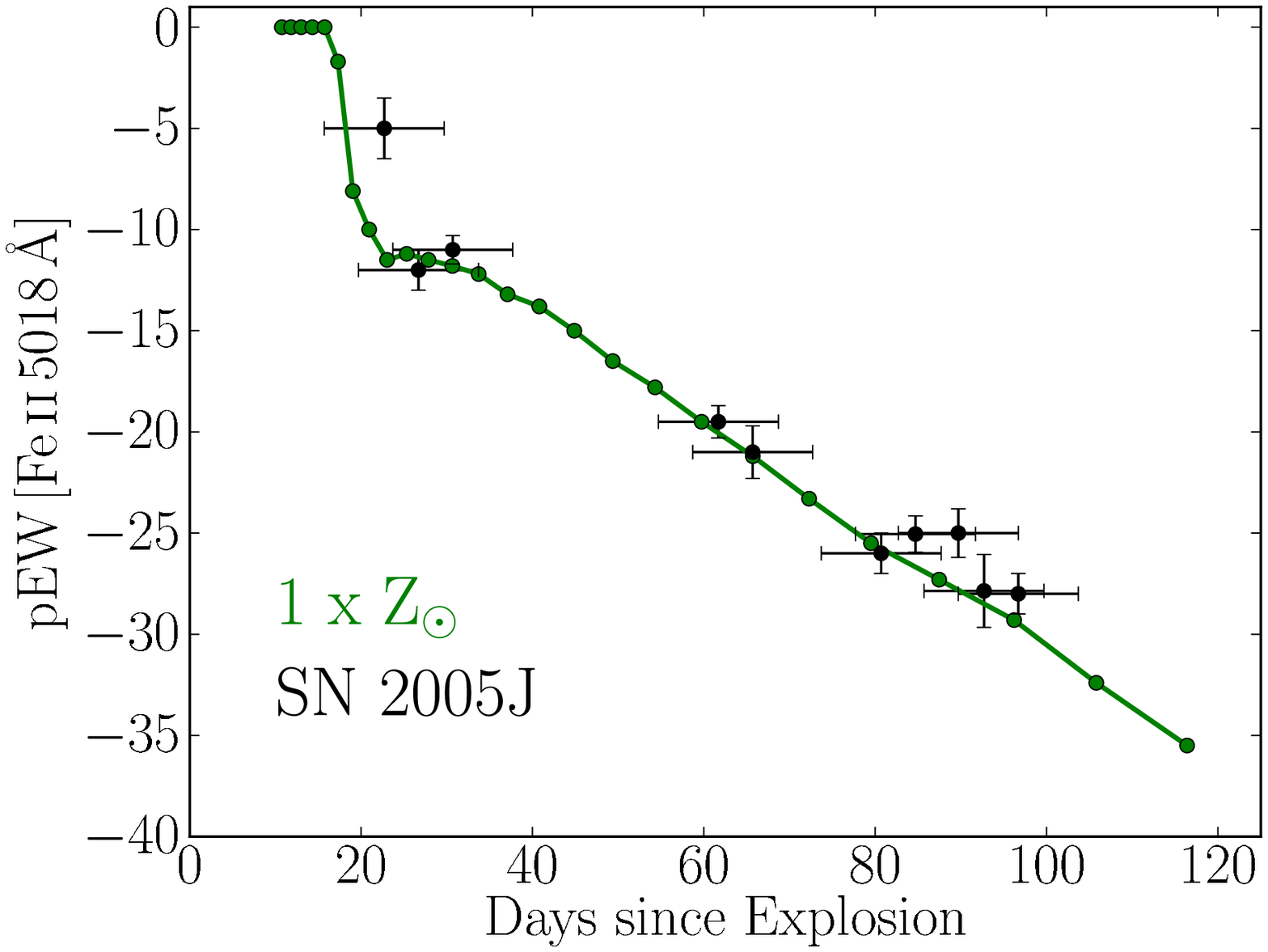,width=5.75cm}
\hspace{0.1cm}
\epsfig{file=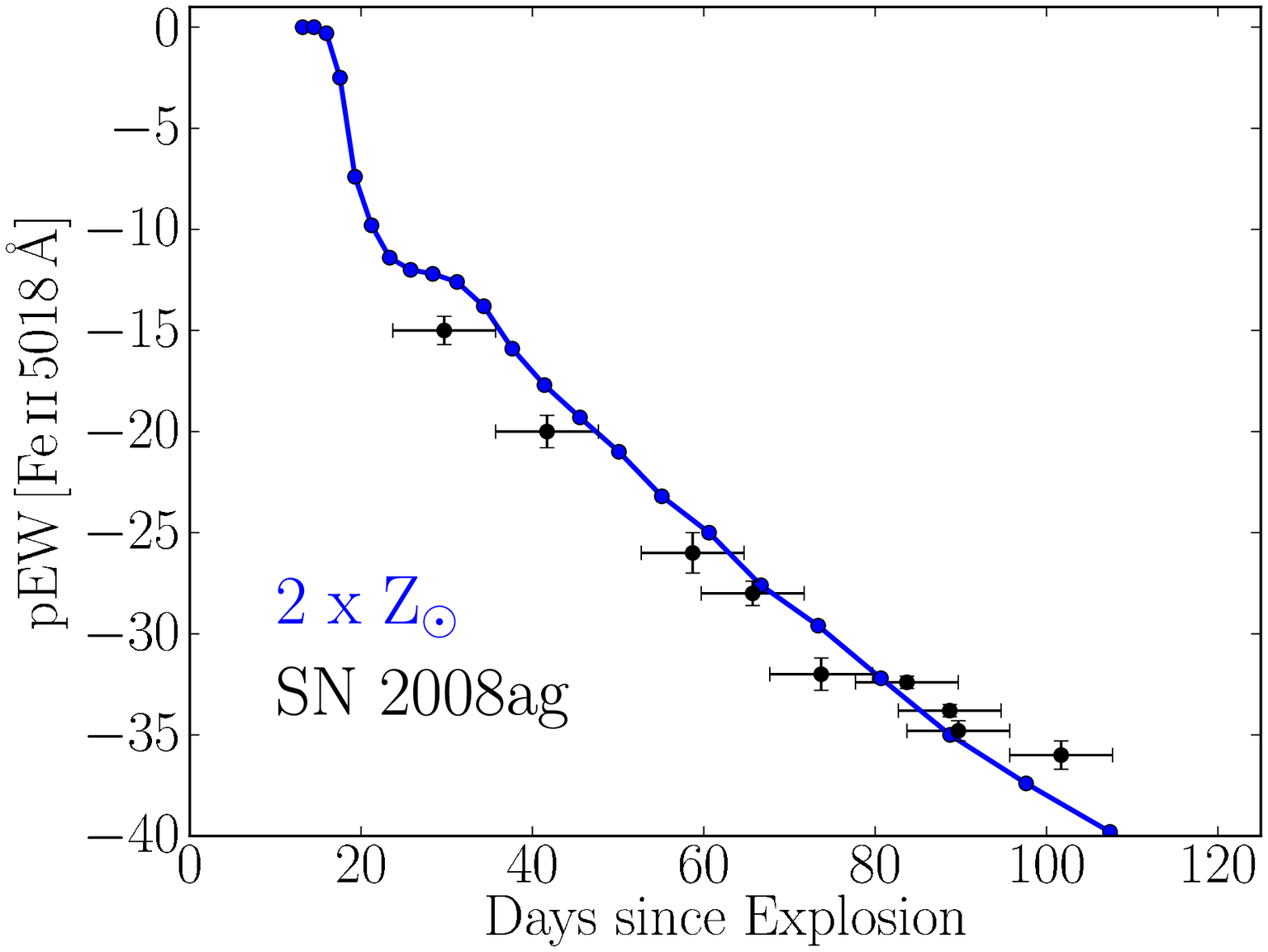,width=5.75cm}
\caption{Comparison of pEWs between observations (SNe 2007il, 2005J, and 2008ag) and
SN II-P models of increasing metallicity (0.4, 1, 2 times solar from left to right).
The same measurement procedure is used on observed and synthetic spectra.
\label{fig_obs_3}
}
\end{figure*}

\section{Comparison to observations}
\label{sect_comp_obs}

Measuring EWs in observed spectra can be difficult.
If we have a radiative-transfer model that matches the observed SN flux
we can also compute the continuum flux and proceed with the rectification
of the spectrum.
Lacking a model, the evaluation of the continuum level
can be very uncertain, in particular short-ward of $\sim$\,5000\,\AA\
during the recombination phase.
In this case, it is customary to measure pEWs,
by calculating the area in absorption bounded by the local flux maxima
around a given line absorption feature.
Such measurements are typically done on features for which
one transition is expected to dominate the absorption, e.g., Fe\two\,5169\,\AA. In reality, this assumption
is not guaranteed and the value of such a measurement may be difficult to interpret.

Although more straightforward than measuring true EWs, measuring pEWs
can still be problematic when line blanketing and overlap are strong, as in model m15z4m2 (Fig.~\ref{fig_spec_seq}).
For H$\alpha$, the presence of overlapping  lines of Si\,\two\ early on and Fe\two\ at the recombination epoch
introduces some ambiguity in the measurement.
The broad Ti\two\ blanketing region is much worse in that respect although it is clear
that the associated absorption  is strongly sensitive to metallicity in our models (see also \citealt{dessart_etal_13b}).
For iron, rather than using the strong Fe\two\,5169\,\AA, which shows signs of overlap, it is more convenient to use
Fe\,\two\,5018\,\AA;  it is bounded by two Fe\two\ lines that make the Fe\two\,5018\,\AA\ absorption dip
easy to measure.

For this work, we use the database of type II SNe from the Carnegie SN Project (CSP; \citealt{hamuy_etal_06})
and other followup programs: the Calan/Tololo SN Survey (CT; \citealt{hamuy_etal_93}),
the Cerro Tololo SN program, the SN Optical and Infrared Survey (SOIRS; \citealt{hamuy_etal_02})
and the Carnegie type II SN Survey (CATS; \citealt{hamuy_etal_09}).
\citet{anderson_etal_14}
have studied their $V$-band light curves to reveal the diversity in fading-rate through the plateau phase,
and thus the presence of intermediate events between plateau and linear SNe II.
Here, we select events that show a rough photometric and spectroscopic compatibility with our four basic
SNe II-P models.
In practice, we exclude SNe that show an appreciable slant in the $V$-band light curve (events with an $s2>$\,1;
see \citealt{anderson_etal_14} for details).
We also exclude events that have a strongly delayed onset of the recombination phase
(suggesting an anomalously slow ejecta cooling, as may arise from a progenitor envelope that is
more extended that normal;  e.g., SN\,2004er; \citealt{iau_04er}; \citealt{anderson_etal_14}).
This rough filtering is necessary in order to isolate spectral diversity associated
with variations in primordial composition rather than with SN energetics or progenitor structure (even if
there may well be a connection between these various progenitor/explosion properties).
Doing this filtering is essential to reduce systematic errors.
In the future, we will need to use a broader sample of progenitor/explosion models at different metallicity
to reflect the type II SN diversity, or produce a tailored model for each SN under study.

\begin{figure*}
\epsfig{file=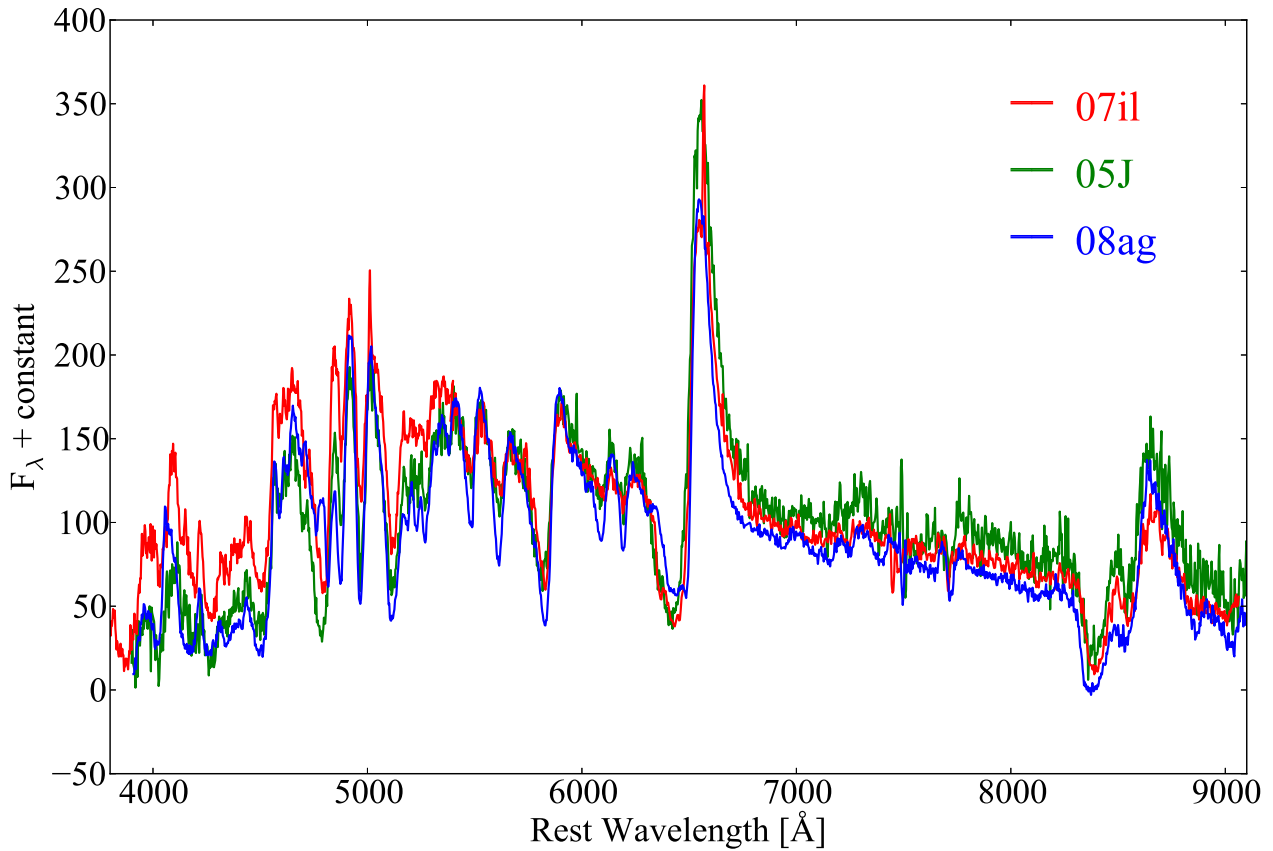,width=19.5cm}
\caption{Montage of spectra for the three SNe II-P shown in Fig.~\ref{fig_obs_3}.
The time is $\sim$\,80\,d after explosion.
\label{fig_obs_spec}
}
\end{figure*}

In Fig.~\ref{fig_obs_3}, we first show the pEW of Fe\two\,5018\,\AA\ for three SNe II-P.
The errors for each datapoint include one on the explosion time. We may set it equal to the
time between detection and prior non-detection, estimate it using {\sc snid} \citep{blondin_tonry_07},
or use an inference from SN modeling (e.g., SN\,1999em; \citealt{DH06_SN1999em}).
The other error is on the pEW, which is estimated from repeating the measurement multiple times.
The pEW in Fe\two\,5018\,\AA\ follows
quite closely the trajectories of models m15z8m3 (0.4$\times$\zsun; 2007il), m15z2m2 (\zsun; 2005J),
and m15z4m2 (2.0$\times$\zsun; 2008ag) --- no SN follows the morphology of model m15z2m3
(0.1$\times$\zsun). Figure~\ref{fig_obs_3} suggests that some observations exhibit  relatively weak or
strong Fe\two\ lines (Fig.~\ref{fig_obs_spec}), in a systematic fashion,
and in that respect reflect our results from sub- to super-solar metallicity models.

\begin{figure*}
\epsfig{file=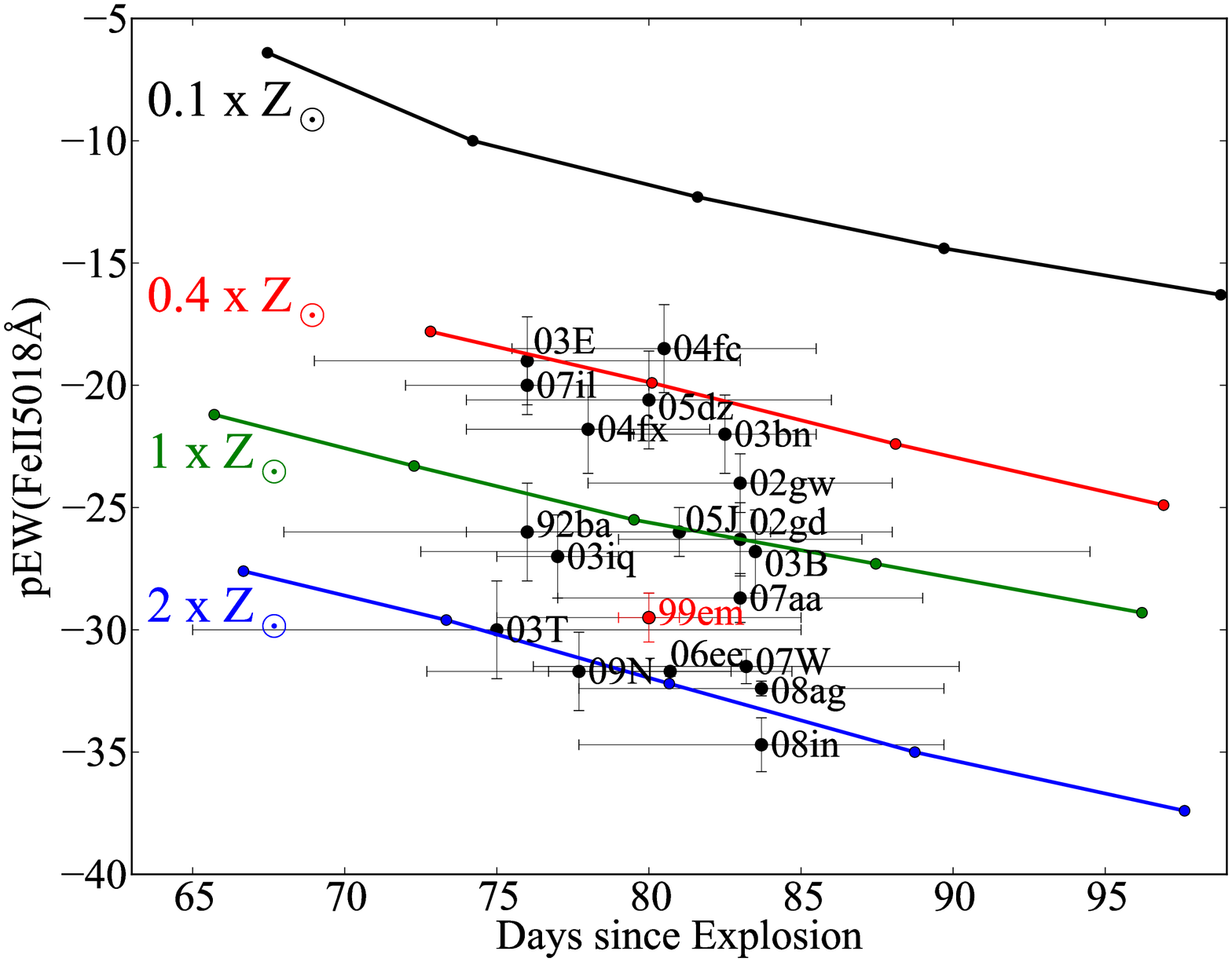,width=18.5cm}
\caption{Distribution of data measurements of pEWs for observed SNe II-P
and the corresponding tracks of theoretical points for models m15z2m3, m15z8m3, m15z2m2, and m15z4m2
(models at 0.1, 0.4, 1, and 2 $\times$ \zsun).
Notice the scarcity of standard SNe II-P in the vicinity of the model at one-tenth solar metallicity.
For comparison, we also add the location of SN\,1999em in this plane (the explosion time for
1999em was inferred by \citet{DH06_SN1999em}, with an error of $\pm$\,1.0\,d).
\label{fig_obs_all}
}
\end{figure*}

When we focus on the recombination phase, around 80\,d after explosion, we find that the
larger sample of type II SNe is well distributed within the tracks of models at
0.4 and 2$\times$\zsun\ (Fig.~\ref{fig_obs_all}).
There is a marked scarcity of SNe II-P at a metallicity below that found in the LMC.
For comparison, we add the location of SN\,1999em, which sits at a slightly super-solar metallicity value.

Progenitor metallicities of core-collapse SNe discovered by means of targeted surveys have been studied
in the past by \citet{anderson_etal_10}.
Using nebular-line analyses of a coincident or nearby H\two\ region, they infer a
$12 + \log[O/H]$ of 8.14 for SN\,2003E, 8.24 for SN\,2002gw, 8.52 for SN\,1992ba, 8.73 for SN\,2003B,
8.61 for SN\,1999em, and 8.64 for SN\,2003T (on the N2 scale of \citealt{pettini_pagel_04}).
These values reflect within the uncertainties the trend we obtain with
our pEW measurements, and thus confirm the results shown in Fig.~\ref{fig_obs_all}.
\citet{anderson_etal_10} do not find any SN II-P at low metallicity, although this may result from using SNe in
high luminosity hosts, which are biased towards high metallicity.
The untargeted survey from the Palomar Transient Factory also have no SN II-P at very low metallicity
\citep{stoll_etal_13}.
In the future, we will perform a more complete analysis of both SNe II-P and host-galaxies/H\two-regions
to compare in detail the metallicity inferred from both techniques. We will also extend the sample to include
SNe II-P from untargeted surveys.

\section{Discussion and conclusions}
\label{sect_conc}

Extending the previous parameter study of \citet{dessart_etal_13b}, we have explored in more detail
the systematic variations of SN II-P spectra with metallicity. Provided one limits the epochs to a few weeks
prior to the end of the plateau phase, the metal lines appearing in SN II-P spectra reflect the
metallicity of the progenitor star, and allow one to place some constraints on the
abundances of oxygen, sodium, titanium, scandium, or iron. This wide range of species spans
from moderate to high atomic mass, and produced by nuclear burning under different conditions
(e.g. steady-state burning in stars versus explosive burning in SNe).

In this paper we have quantified the variations of the line-absorption EW with metallicity in several strong and/or isolated
lines including O\one\,7777\,\AA, Na\one\,D, Ti\two\ absorption at 4200-4500\,\AA, Fe\two\,5018,
Fe\two\,5169\,\AA, as well as the Ca\,\two\,7300\,\AA\ emission doublet.
More lines could be studied but this diverse set is good for
the proof-of-concept discussion.
We also avoid using the UV range due to the low fluxes, saturated and badly blended lines,
and the sensitivity to parameters other than metallicity.
We find that in all models, the evolution through the plateau phase leads to a systematic increase
in the magnitude of these EWs. H$\alpha$ shows little sensitivity to metallicity, while Na\one\,D becomes
markedly weaker only when the metallicity  is decreased to one-tenth solar. However, O\one\,7777\,\AA,
Fe\two\,5169\,\AA, and Fe\two\,5018\,\AA\ exhibit a systematic trend at all times that correlates
with the metallicity of the model.
A good time for analysis is during the recombination phase, when metal-line blanketing is strong,
but before the end of the plateau to avoid any pollution at the photosphere, as may occur through the mixing
of species from the helium core into the H-rich envelope.

We have confronted the models to a selection of SNe II-P from the CSP and former followup programs
\citep{hamuy_etal_06,anderson_etal_14}.
Calculating pseudo-EWs on both observations and models at the recombination epoch, we find
they exhibit the same behavior from the early-photospheric phase until the end of the plateau
(e.g., increasing EW magnitude in Fe\two\,5018\,\AA).  Comparing the measurements
at $\sim$\,80\,d after explosion,  we find
that all selected SNe fall within the pEW limits set by the 0.4 and 2$\times$\,\zsun\ models. No SN II-P
in our sample matches the weak metal-line strengths of our model at \zsun/10. Since the CSP observations
are probably representative of type II SNe, we speculate that we are yet to observe a SN II-P at SMC
metallicity in the local Universe.

Our metallicity measurements compare favorably with nebular-line analyses.
Such studies, based on both targeted   \citep{anderson_etal_10} and untargeted \citep{stoll_etal_13}
surveys confirm the scarcity of SNe II-P at low metallicity.
The absence of low metallicity SNe in our sample most likely reflects the paucity of very low metallicity systems
in the local universe. The absence of low metallicity hosts is also seen in the analysis of
core collapse SNe out to $z\sim$\,0.2 by \citet{kelly_etal_14}.
Since stellar evolution depends on metallicity it is important to study SN at low metallicities.
Extreme rotation rates on the main sequence, for example,  can prevent a star, if it has an initial mass of $\gtrsim$\,20\,\msun,
from evolving to the red and from exploding as a RSG  (see, e.g., \citealt{brott_etal_11}). Irrespective of rotation, stars
in the mass range 10--20\,\msun\ are expected to explode in a RSG phase at low metallicity, and thus should be seen.

In addition to providing important constraints on the SN and its progenitor, quantitative spectroscopy
of SNe can also be used for determining the evolution of  metallicity  with redshift,  and for revealing
the metallicity distribution with galactocentric radius. The high luminosity of SNe makes them
ideal substitutes to stars (see, e.g., \citealt{kudritzki_etal_12}), but may also offer an alternative
to nebular-line analyses \citep{osterbrock_89,kewley_dopita_02,pettini_pagel_04}.
At very large distances, we will need super-luminous SNe resulting from the pair-production instability in
a super-massive RSG-star, since their plateau luminosities are predicted to be on the order of 10$^{10}$
rather than a few 10$^8$\,\lsun\ \citep{kasen_etal_11,dessart_etal_13}.
Unfortunately, such super-luminous SN II-P events are yet to be discovered.

With {\it VLT-FORS}, a SN II-P of 14th magnitude at 10\,Mpc requires a 0.7\,s exposure to yield a
S/N of 30 per pixel (grism 150I). At 100\,Mpc, this exposure time is 70\,s. Using a Hubble constant
of 70\,\kms\,Mpc$^{-1}$, that same SN at a redshift of 0.1 would require an exposure of about 30\,min
for the same setup. With the future extremely-large telescopes, going up to redshift one and beyond
may be possible. In practice, to reduce systematic errors from the modeling and to circumvent
inaccurate pseudo-equivalent widths measurements, it will be desirable to perform
detailed modeling of each SN II-P under study.

\section*{Acknowledgments}

LD acknowledges financial support from the Unit\'e Mixte Internationale 3386 (Laboratoire Franco-Chilien
d'Astronomie, CNRS/INSU France, Universidad de Chile),  from the European Community through
the International Re-integration Grant PIRG04-GA-2008-239184, and from the ``Agence Nationale de la Recherche"
through the grant ANR-2011-Blanc-SIMI-5-6-007-01.
C.P.G. acknowledges support from CONICYT-AGCI PhD studentship.
C.P.G., M.H., J.P.A, and F.L acknowledge support from the Millennium Center for Supernova Science through
grant P10Ð064-F and the Millennium Institute for Astrophysics grant IC120009, funded
by Programa Iniciativa Cientifica Milenio.
DJH acknowledges support from STScI theory grant
HST-AR-12640.01, and NASA theory grants NNX10AC80G and NNX14AB41G.
This work was granted access to the HPC resources of CINES (France) under the allocation c2013046608
made by GENCI (Grand Equipement National de Calcul Intensif).
G.~F. acknowledges financial
support by Grant-in-Aid for Scientific Research for Young Scientists
(23740175). This material is based upon work supported by NSF under
grants AST--0306969, AST-0908886, AST--0607438, and AST-1008343.
M.~S. acknowledges the generous support provided by the
Danish Agency for Science and Technology and Innovation through a
Sapere Aude Level 2 grant.

\label{lastpage}

\end{document}